\documentclass[12pt,preprint]{aastex}

\shorttitle{Scattered-Light Imaging of Six Young Stars}
\shortauthors{Doering et al.}

\begin{document}


\title{HD 97048's Circumstellar Environment as Revealed by a {\it HST}/ACS Coronagraphic 
	Study of Disk Candidate Stars}



\author{R.~L. Doering\altaffilmark{1,2},
M. Meixner\altaffilmark{2},
S.~T. Holfeltz\altaffilmark{2},
J.~E. Krist\altaffilmark{3},
D.~R. Ardila\altaffilmark{4},
I. Kamp\altaffilmark{5},
M.~C. Clampin\altaffilmark{6},
\& S.~H. Lubow\altaffilmark{2}} 

\altaffiltext{1}{Department of Physics, University of Illinois, 1110 West Green Street, Urbana, IL 61801.}
\altaffiltext{2}{Space Telescope Science Institute, 3700 San Martin Drive, Baltimore, MD 21218.}
\altaffiltext{3}{Jet Propulsion Laboratory, 4800 Oak Grove Drive, Pasadena, CA 91109.}
\altaffiltext{4}{{\it Spitzer} Science Center, IPAC, MS 220-6, California Institute of Technology,
Pasadena, CA 91125.}
\altaffiltext{5}{Space Telescope Division of ESA, Space Telescope Science Institute, 3700 San Martin
Drive, Baltimore, MD 21218.}
\altaffiltext{6}{NASA Goddard Space Flight Center, Code 681, Greenbelt, MD 20771.}


\begin{abstract}
We present the results of a coronagraphic scattered-light imaging survey of six young disk candidate stars using the {\it Hubble Space Telescope} Advanced Camera for Surveys.  The observations made use of the 1\farcs8 occulting spot through the F606W (broad $V$) filter.  Circumstellar material was imaged around HD 97048, a Herbig Ae/Be star located in the Chamaeleon I dark cloud at a distance of 180 pc.  The material is seen between $\sim$2$\arcsec$ (360 AU) and $\sim$4$\arcsec$ (720 AU) from the star in all directions.  A $V$-band azimuthally-averaged radial surface brightness profile peaks at r $=$ 2$\arcsec$ with a value of 19.6 $\pm$ 0.2 mag arcsec$^{-2}$ and smoothly decreases with projected distance from the star as {\it I} $\propto$ {\it r}$^{-3.3 \pm 0.5}$.  An integrated flux of 16.8 $\pm$ 0.1 mag is measured between 2$\arcsec$ and 4$\arcsec$, corresponding to a scattered-light fractional luminosity lower limit of L$_{sca}$/L$_{*}$ $>$ $8.4 \times 10^{-4}$.  Filamentary structure resembling spiral arms similar to that seen in Herbig Ae/Be disks is observed.  Such structure has been attributed to the influence of orbiting planets or stellar encounters.  Average surface brightness upper limits are determined for the five non-detections: HD 34282, HD 139450, HD 158643, HD 159492, and HD 195627.  Possible reasons for the non-detections are disks that are too faint or disks hidden by the occulter.
\end{abstract}




\section{Introduction}

The study of dusty circumstellar disks around young stars informs our understanding of the planet formation process and the development of planetary systems including our solar system.  In the first few million years of a star's life, planets may form in circumstellar disks through a poorly understood process of grain growth and gravitational instabilities followed, in the case of giant planets, by a gas accretion phase \citep[e.g.][]{lis95}.  As a young star evolves towards the main sequence, its protoplanetary disk changes due to accretion of material onto the star, outflows, radiation pressure, and planet formation.  As a protoplanetary disk ages, these processes cause dust emission to decrease as small grains coagulate into larger bodies, or are removed from the system \citep[e.g.][]{bec96, zuc01}.  Evolution continues even after the epoch of planet formation as collisions between residual planetesimals replenish dust in debris disks \citep[e.g.][]{hab01}.  

High-resolution observations reveal disk morphologies that are used to break degeneracies that plague analyses based on spectral energy distributions (SEDs) alone.  The imaging of circumstellar disks has been greatly enhanced by the instruments of the {\it Hubble Space Telescope} ({\it HST}).  In particular, debris disks and Herbig Ae/Be disks have been resolved in scattered light through coronagraphic observations with NICMOS \citep[e.g.][]{sch99, sch05}, STIS \citep[e.g.][]{gra00, gra01, gra05}, and ACS \citep[e.g.][]{cla03, ard04, kri05, kal06}.  The stars with resolved disks exhibit a wide range of spectral types (B9.5 - M1) and reside in both pre-main-sequence and main-sequence stages of stellar evolution.  Many of the disks show warps, clearings, and spiral structure that may indicate the presence of planets.  Scattered-light imaging reveals the disk independent of its temperature, thereby eliminating the need to detect thermal emission.  This is particularly important in resolving the outer regions of a disk where the dust grains may be too cold to detect in mid- to far-IR emission, and the resolution provided by current (sub)millimeter observations is often not adequate.

Intermediate mass stars ($\sim$2 -- 10 $M_{\sun}$) are particularly accessible candidates for disk imaging.  They are bright enough to have been identified as IR excess sources by {\it IRAS} and/or {\it ISO} mid- to far-IR photometry.  Complementary ground-based data are frequently available, or can be readily obtained with follow-up observations.  Most well-studied debris disk systems in the intermediate mass range (e.g. Vega, $\beta$ Pictoris) are thought to be preceded by a Herbig Ae/Be phase.  However, the transition from a primordial protoplanetary dust disk to one in which the circumstellar material is replenished via collisions between orbiting bodies is not well understood.  Additionally, intermediate mass stars provide a link between disk evolution in low mass, post-T Tauri stars, and high mass pre-main sequence stars such as those found in ultra-compact HII regions \citep{wat98}.

Among the Herbig Ae/Be stars, the circumstellar dust may not be confined to a disk-like geometry.  Observations of these systems have also been interpreted as dust distributed in envelopes instead of disks, or as a compact disk component embedded within a spherical envelope component \citep[e.g.][]{mir99, mil01, eli04}.  Indeed, the high-resolution {\it HST} observations of these systems have revealed prominent non-disk features.  This is consistent with an evolutionary scenario where an infalling envelope serves as a reservoir for the disk, and eventually dissipates as the material is used for planet formation, captured in a debris disk, and escapes the system.    

We carried out a coronagraphic imaging study of three Herbig Ae/Be stars and three debris disk candidate stars using the {\it Hubble Space Telescope} Advanced Camera for Surveys in order to spatially resolve the distribution of circumstellar material around these young stars, and contribute to the growing database of disk systems.  The source selection, coronagraphic observations, and data reduction procedure are discussed in $\S$2.  The imaging results and a discussion of the data are presented in $\S$3 and $\S$4, respectively. The main conclusions of the present work are summarized in $\S$5.

\section{Source Selection, Observations, and Data Processing}

\subsection{Source Selection} \label{targets}

Expanding the number of spatially resolved young disk systems was a primary goal of this study, and thus, our six sources were selected from a pool of candidate stars found in the literature that had not been previously imaged in scattered light with the ACS coronagraph.  All sources have been identified as either Herbig Ae/Be stars or debris disk candidate systems with spectral types ranging from A -- G.  The classification of HD 158643 (51 Oph) as a Herbig Ae/Be star is not very well established \citep[e.g.][]{van01, mee01}, but it is considered one for the purposes of this paper.  Their measured IR excesses indicate the presence of circumstellar material and have values that are within the range of L$_{IR}$/L$_{*}$ values of disks previously resolved with ACS ($\ga$ 10$^{-5}$).  Published multi-wavelength photometry that yields well-constrained spectral energy distributions is available for each source.   As a group, the stars have been observed at mid-IR wavelengths with {\it IRAS}, {\it ISO}, the {\it Spitzer Space Telescope}, and various ground-based imagers and spectrographs.  Circumstellar gas has been probed in most of the sources through millimeter CO observations or optical/infrared spectroscopy.  The infrared and millimeter observations are generally interpreted as emission arising from a circumstellar disk in each of these systems.  Stellar proximity was also a selection criterion since a non-detection due to circumstellar material hidden behind the occulter becomes more likely with increasing distance.  Thus, the selected sources are all located at  d $<$ 200 pc based on {\it Hipparcos} parallax measurements.  However, \citet{pie03} have suggested a revised distance of 400 pc for HD 34282, well above the upper limit set by the {\it Hipparcos} uncertainty.  Properties of our disk candidate stars and corresponding PSF reference stars are listed in Tables 1 and 2, respectively.

The reported infrared excesses and SED profiles indicate that our sample includes disk systems from both the optically-thin and optically-thick regimes.  The Herbig Ae/Be stars HD 34282 and HD 97048 are likely surrounded by optically-thick disks.  HD 158643 may be an intermediate case with a marginally optically-thin disk in which most of the material is close to the star.  The three closest systems, HD 139450, HD 159492, and HD 195627, are considered optically-thin main-sequence debris disk stars.

\subsection{Observations and Data Processing} \label{obs}

The stars were observed with the {\it Hubble Space Telescope} ({\it HST}) Advanced Camera for Surveys (ACS) High Resolution Channel (HRC) during Cycle 13.  The HRC has a pixel scale of $\sim$0\farcs025 pixel$^{-1}$ and a point-spread function (PSF) FWHM of 63 mas at $V$.  As this observational program was a search for circumstellar disks, the targets were imaged through a single filter, F606W (broad $V$).   The 1\farcs8 occulting spot was used for all coronagraphic exposures.  Observational parameters for all targets are listed in Table 3.  In addition to the occulted exposures, direct, non-coronagraphic, exposures were taken of each star in order to obtain photometry.  Exposure times were determined from previous ACS coronagraphy programs (e.g. 9295, 10330) and the ACS Exposure Time Calculators.\footnote{ACS Exposure Time Calculators, available at http://www.stsci.edu/hst/acs/software/etcs/ETC\_page.html}  

Each disk candidate star was assigned a PSF reference star for subtraction.  The PSF reference stars were selected based on their proximity on the sky to the disk candidate, $V$ magnitude brighter than the disk candidate, and similar (\bv) color and spectral type.  Disk candidate stars and the corresponding PSF reference stars were observed on consecutive orbits.  The recommended observing strategy presented in the ACS Instrument Handbook\footnote{ACS Instrument Handbook (Gonzaga et al. 2005), available at http://www.stsci.edu/hst/acs/documents.} was followed: the disk candidate stars were observed with direct exposures followed by the coronagraphic exposures, whereas the PSF reference stars were observed in the coronagraphic mode first, followed by the direct exposures.  This was done so the coronagraph, which can change position slightly when cycled, was not reconfigured between the observations of the disk candidate and corresponding PSF reference star. 

All images were initially processed with the {\it HST} pipeline.  A pixel area map correction was applied to each direct exposure image according to the ACS Data Handbook\footnote{ACS Data Handbook (Pavlovsky et al. 2004), available at http://www.stsci.edu/hst/acs/documents.} in order to perform point-source photometry.  This correction accounts for the geometric distortion present in flat-fielded {\it HST}/ACS images by compensating for pixel area distortion.  Aperture photometry was performed on each disk candidate and PSF reference target using a 5$\farcs$5 (220 pixel) radius.  Several photometric observations were saturated.  However, an HRC gain of 4 e$^{-}$ per data unit ensures that the flux from saturated pixels will overflow into surrounding pixels so data is not lost.\footnote{ACS Instrument Science Report ISR 04-01 (Gilliland 2004), available at http://www.stsci.edu/hst/acs/documents/isrs.}  The aperture was large enough to capture all the flux in each image.  The transformation from counts s$^{-1}$ to standard $V$ magnitudes was carried out with the STSDAS synthetic photometry package SYNPHOT assuming the spectral types given in Tables 1 and 2.  The photometric measurements are shown in Table 4.  The photometry was used to normalize the coronagraphic PSF reference star exposures before subtraction from the disk candidate star images.  Examples of occulted disk candidate and PSF reference star images prior to subtraction are shown in Figure 1.

The PSF reference images were iteratively shifted using a cubic convolution interpolation and subtracted from the disk candidate images until the residual scattered-light patterns were visibly minimized.  The final alignments are good to within $\pm$0.05 pixels. After the best shifts were determined, the residual images were corrected for geometric distortion.  This correction is carried out using a set of distortion coefficients in a function that outputs an undistorted 0\farcs025 pixel$^{-1}$ image to a 1208 $\times$ 1208 pixel array.  The geometric distortion correction was performed after the registration and subtraction to avoid errors which may be introduced by the routine's interpolation.  The undistorted residual images were rebinned (2 $\times$ 2) in an effort to enhance the visibility of any circumstellar material.  The residual images are dominated by subtraction errors out to a radius of $\sim$2$\arcsec$ from the occulted star's position.

Surface brightnesses were converted to Jy arcsec$^{-2}$ and mag arcsec$^{-2}$ using the CALCPHOT task in the SYNPHOT package.  Using the calibrated PHOTFLAM value, we derived the following set of surface brightness conversions for the A0 stars:  1.49 $\times$ 10$^{-7}$ Jy arcsec$^{-2}$ per counts s$^{-1}$ arcsec$^{-2}$ and SB(mag arcsec$^{-2}$) $=$ -2.5log[SB(mJy arcsec$^{-2}$)] $+$ 16.4.  This assumes the counts s$^{-1}$ value has been sky subtracted and includes the 47.5\% throughput correction for coronagraphic observations discussed in the ACS Data Handbook.

\section{Results}

\subsection{Detection of Circumstellar Material Around HD 97048} \label{disk results}

The PSF-subtracted images revealed resolved circumstellar material around one of the sources in our survey, the Herbig Ae/Be star HD 97048 (Figure 2).   The material extends to a radial distance of $\sim$4$\arcsec$ in almost all directions, or $\sim$720 AU given the assumed distance of 180 pc.  The inner radius cannot be determined from our {\it HST}/ACS coronagraphic image due to the subtraction residuals that dominate the region within an angular distance of $\sim$2$\arcsec$ from the star. The image is not contaminated by background objects, though a few negative regions appear in the field shown in Figure 2 because of foreground/background objects in the PSF reference star field.

The circumstellar nebulosity does not have a spatially uniform azimuthal distribution, but instead appears to have structure.  Several features in HD 97048's circumstellar material are indicated with arrows in Figure 3.  Spiral-like structure is especially prominent on the material's north side at a projected distance of $\sim$3$\arcsec$ with a subtraction artifact passing through it (labeled A in Figure 3).  We estimate a signal-to-noise ratio (S/N) of 6.1 for this feature.  A faint extension (Figure 3, B) that goes beyond 4$\arcsec$ with a S/N of 3.7 is observed on the northeast side.  It appears to merge with the surrounding reflection nebulosity on the sky.  Clumpy features (S/N $=$ 3.1) are seen in the system's southern region (Figure 3, C).  Faint filamentary structure also appears in the southwest (Figure 3, D) and northwest regions (Figure 3, E) with S/N of 2.9 and 2.4, respectively.  It is not known if these filaments are associated with a disk or if they are part of a circumstellar envelope.   

We measure a $V$-band azimuthally-averaged peak surface brightness of 19.6 $\pm$ 0.2 mag arcsec$^{-2}$ at a projected distance of {\it r} $=$ 2$\arcsec$ (the approximate outer edge of the region dominated by subtraction residuals).  The radial surface brightness profile of the circumstellar material from {\it r} $=$ 2$\arcsec$ to {\it r} $=$ 7$\arcsec$ appears smooth and decreases nearly monotonically (Figure 4).  A power-law fit to the profile yields {\it I} $=$ [(5.6 $\pm$ 2.9) $\times$ 10$^{-4}$]{\it r}$^{-3.3 \pm 0.5}$ $+$ (3.2 $\pm$ 1.1) $\times$ 10$^{-6}$  (units of {\it I} are Jy arcsec$^{-2}$ and {\it r} is in arcsec).  The integrated flux from {\it r} $=$ 2$\arcsec$ to {\it r} $=$ 4$\arcsec$ is 16.8 $\pm$ 0.1 mag.  This flux corresponds to a scattered-light fractional luminosity of L$_{sca}$/L$_{*}$ $>$ $8.4 \times 10^{-4}$, smaller than HD 97048's IR fractional excess by a factor of $\la$ 475.  The discrepancy between the scattering luminosity and IR excess is expected given that much of the scattering is done in the occulted region, and if the material is distributed in an optically thick disk, most of the mass resides in the dark midplane.  HD 97048's circumstellar material properties determined from our scattered-light image are summarized in Table 5.

Previous observations of HD 97048 have provided evidence for both disk and envelope dust distribution components.  \citet{the86} concluded based on a photometric and spectroscopic study that HD 97048's infrared excess is due to dust in a flattened shell structure or pole-on disk.  \citet{dav91} report that the observed reddening of {\it E(\bv)} $=$ 0.36 is mostly intracloud and the amount of circumstellar reddening is actually negligible, consistent with a distribution seen pole-on. The unresolved SEST 1.3 mm map and flux measurement indicates dust distributed in a disk-like geometry \citep{hen93, hen98}. Spatially resolved nanodiamond features at 3.43 and 3.53 $\mu$m consistent with emission from the inner region ($<$ 15 AU) of a pole-on disk have been observed \citep{hab04}.  Additionally, \citet{ack06} observed double-peaked [OI] emission around HD 97048 that they attribute to a rotating disk with a position angle of 160 $\pm$ 19$\degr$ east of north.

The mid-IR spectral regime has been used to probe both components' structure and composition.  Mid-IR extended emission has been detected on scales of 5 -- 10$\arcsec$ and modeled as an envelope of transiently heated very small grains and PAHs \citep{pru94, sie00}.  More recent high-resolution mid-IR spectroscopic observations by \citet{van04} have measured spatially resolved 10 $\mu$m emission (continuum and {\it UIR} bands) within a few hundred AU of the star.  The authors propose that this emission originates in a flared disk.  The 8.6 $\mu$m imaging of HD 97048 by Lagage et al. (2006) revealed spatially extended emission in the system's inner region ($<$ 2$\farcs$1) as well.  They model their observational results as PAH emission from the surface layer of a vertically optically thick flared disk with an inclination of 42.8$\degr$ from pole-on.

Based on the observational evidence, in particular the mid-IR studies, it seems likely that HD 97048's circumstellar environment consists of a flared disk plus a diffuse remnant envelope.  There is little spatial overlap between our scattered-light image and the previous mid-IR observations.  Instead, our image probes an intermediate region approximately bounded by the extended envelope emission detected by \citet{pru94} and \citet{sie00} and the resolved inner disk region \citep{van04, lag06}.  Our scattered-light image does not appear to contradict the mid-IR results and could be interpreted as the outer region of an inclined disk.  However, we cannot rule out a contribution to our image from a circumstellar envelope.

\subsection{Non-detections} \label{no disks}

PSF-subtracted images of the other program stars (HD 34282, HD 139450, HD 158643, HD 159492, and HD 195627) do not reveal the presence of circumstellar material seen in scattered light in these systems.  An example of a non-detection among the Herbig Ae/Be stars (HD 34282) is shown in Figure 5.  No material is observed beyond the subtraction residual region.  An azimuthally-averaged radial surface brightness profile of HD 34282 falls off rapidly from 21.4 to 22.8 mag arcsec$^{-2}$ between r $=$ 2$\arcsec$ to 3$\farcs$5, compared to the more gradual fall-off measured in the HD 97048 system (Figure 6). 

PSF mismatches of varying degree are observed in the subtracted images for HD 139450, HD 158643, HD 159492, and HD 195627.  The mismatch results in the presence of an artificial halo around the subtracted target.  These mismatches may be caused by Optical Telescope Assembly (OTA) breathing or a slight color difference between the disk candidate and PSF reference stars.  In order to quantify the non-detections, we report upper limits to the surface brightness in the region where the presence of a disk might be expected and sky brightness upper limits.  The disk region was defined as an annulus centered on the spot between radii of 2$\arcsec$ and 4$\arcsec$ (80 and 160 pixels).  The sky measurements were made in an annular region between radii of 5$\farcs$6 and 6$\farcs$5 (224 and 260 pixels), also centered on the spot.  Table 6 lists the disk surface brightness 3$\sigma$ upper limits and the sky brightness 3$\sigma$ upper limits for the five non-detections.

\section{Discussion}

The number of Herbig Ae/Be stars with circumstellar material resolved in scattered light remains relatively small.  In addition to HD 97048, material around HD 100546, AB Aurigae (AB Aur), HD 163296, and HD 141569A has been resolved by {\it HST} coronagraphy.   HD 141569A may be a transitional object between the pre-main sequence protoplanetay disk and main-sequence debris disk phases  \citep{cla03} and has been included as an upper evolutionary bound.  Table 7 lists stellar and circumstellar material properties for these previously-imaged systems, HD 97048, and the two Herbig Ae/Be non-detections from this study (HD 34282 and HD 158643).

Each of the previously-imaged Herbig Ae/Be systems show complex morphology with structures such as spiral arms, dark and bright lanes, and gaps.  These structures have been generally attributed to the effects of orbiting planets or encounters with nearby stars.  The circumstellar material around HD 97048 is currently the most distant to be resolved in scattered light with {\it HST} coronagraphic observations.  Therefore, the material's features will be more difficult to identify than its closer counterparts.  However, a comparison between the previously-imaged systems and HD 97048 will permit a qualitative analysis of our data in terms of the larger pool of resolved disk, or disk plus envelope, Herbig Ae/Be systems.  Spiral-like structure is a trait that HD 97048 shares with AB Aur, HD 100546, and HD 141569A \citep{gra99, gra01, cla03}, though it is less prominent than that observed in the other systems.  A large-scale clearing, such as the one found around HD 141569A, is not observed in the HD 97048 circumstellar material.  This may indicate that such a clearing is not present in the HD 97048 system, or we are unable to resolve it given the star's distance.  Like AB Aur, the surface brightness of HD 97048's circumstellar material decreases nearly monotonically with radius without the suggestion of large gaps, though the radial falloff in HD 97048 is steeper than that observed around AB Aur \citep[{\it I} $\propto$ {\it r}$^{-2}$;][]{gra99}. However, HD 97048's surface brightness falloff is in good agreement with the value measured for HD 163296's outer disk \citep[{\it I} $\propto$ {\it r}$^{-3.5}$ beyond 370 AU;][]{gra00} and HD 100546's disk region \citep[{\it I} $\propto$ {\it r}$^{-3.1}$ within 5$\arcsec$;][]{gra01}.  The observations of HD 97048 taken in total suggest that it belongs to a group of Herbig Ae/Be stars that have circumstellar environments consisting of a disk plus an extended envelope.  HD 100546, HD 163296, and AB Aur are also likely members of this group.

Despite evidence that indicated the presence of disks, or more generally, circumstellar material, around all of our targets, there were 5 non-detections. One explanation is that these targets are surrounded by disks, but their projected size is small enough to be completely hidden behind the occulting spot or lost in the PSF-subtraction residuals.  Given its revised distance of 400 pc, this may be the case for HD 34282.    The CO $J\!=\!2\!\rightarrow\!1$ line emission, 1.3 mm continuum, and 3.4 mm continuum observations of \citet{pie03} indicate a CO disk in Keplerian rotation with a derived outer radius of 835 AU.  At that distance, the disk's outer radius would be near the inner edge of the ACS coronagraph's measurable region ($\sim$2$\arcsec$).  \citet{den05} observed a double-peaked $^{12}$CO $J\!=\!3\!\rightarrow\!2$ line profile and estimated a disk outer radius of 360 AU at 400 pc, further suggesting that the disk is inaccessible to the ACS coronagraph.

A disk hidden behind the occulter may account for the HD 158643 (51 Oph) non-detection as well.  This source's SED has a significant near- to mid-IR excess, but then drops off longwards of ~25 $\mu$m, suggesting an absence of cold grains \citep[e.g.][]{jay01, lei04}.  \citet{thi05} reported the detection of CO bandhead emission, indicating the presence of hot (2000 -- 4000 K) and dense molecular gas in the system.  The authors model the emission as originating from a nearly edge-on dust-poor gas disk which is within an AU of the star and undergoes Keplerian rotation.  \citet{sch99} used 51 Oph as a PSF reference star for NICMOS coronagraphic imaging of HR 4796A, implying they saw no evidence of  extended circumstellar material in their observations of this source.  Also, \citet{sch04} did not detect a disk with ground-based near-IR coronagraphic imaging.  Taken together, the observations suggest a very compact disk surrounding 51 Oph that is blocked by the occulter in coronagraphic imaging.

The debris disk candidate non-detections (HD 139450, HD 159492, HD 195627) could also be the result of disks hidden by the occulter.  \citet{kal06} introduced the classification of debris disks into two categories based on the radial dust distribution: 20 -- 30 AU wide narrow belts and more extended disks $>$ 50 AU in width.  The narrow belt systems may be particularly susceptible to total coverage by the coronagraphic mask.  Another possibility is that a disk's surface brightness may be too faint to be detected by ACS.  Our closest targets, HD 159492 and HD 195627, have estimated ages of 200 Myr and IR optical depths (L$_{IR}$/L$_{*}$) of $1 \times 10^{-4}$ \citep{zuc04}, at the lower end of the range for detected disks. The nearby Fomalhaut ring-like disk system ({\it d} $=$ 7.7 pc) has L$_{IR}$/L$_{*}$ $=$ $5 \times 10^{-5}$, the lowest IR optical depth of any debris disk resolved with the ACS coronagraph thus far \citep{kal05}.  The HD 139664 disk system has an IR optical depth of $0.9 \times 10^{-4}$ \citep{kal06}, comparable to HD 159492 and HD 195627.  However, HD 139664 is seen nearly edge-on.  Therefore, disks around HD 159492 and HD 195627 may have dissipated to a level that is not detectable with current instruments.  If the circumstellar material is present in a more diffuse, low-surface brightness distribution such as an envelope, or a near face-on oriented disk, it may fall below the detection threshold as well.  We also note HD 139450's uncertain optical-{\it IRAS} association as another possible reason for not detecting circumstellar material around this source \citep{syl00}.  Properties of our debris disk candidate non-detections are compared to a subset of debris disks detected with ACS in Table 8.

\section{Summary}

The scattered-light coronagraphic images reveal circumstellar material around the Herbig Ae/Be star HD 97048 extending to $\sim$4$\arcsec$.  The material has a $V$-band azimuthally-averaged peak surface brightness of 19.6 $\pm$ 0.2 mag arcsec$^{-2}$, with a radial profile that decreases monotonically as  {\it I} $\propto$ {\it r}$^{-3.3 \pm 0.5}$ .  We measure an integrated flux of 16.8 $\pm$ 0.1 mag, yielding a scattered-light fractional luminosity lower limit of $>$ $8.4 \times 10^{-4}$.  The material has azimuthal variations similar to spiral-like structure observed in Herbig Ae/Be disks coronagraphically imaged in scattered-light.  One intriguing possibility is that such features arise from the presence of perturbing bodies in the system.  Multiband scattered-light coronagraphic imaging of HD 97048 is needed to permit a more detailed analysis and further our understanding of the system's circumstellar dust properties.  We also report five non-detections: HD 34282, HD 139450, HD 158643 (51 Oph), HD 159492, and HD 195627.  Possible explanations for these non-detections include surface brightness levels below the detection threshold of the ACS HRC, or a disk that is completely hidden by the coronagraph's occulting spot.

\acknowledgments

This work was supported by NASA/STScI GO-10425.01 and NASA grant NAG5-12595.  This research has made use of the SIMBAD database, operated at CDS, Strasbourg, France.  We thank E. Pantin for useful discussions about the HD 97048 VISIR image.



\clearpage


\begin{figure}
\plottwo{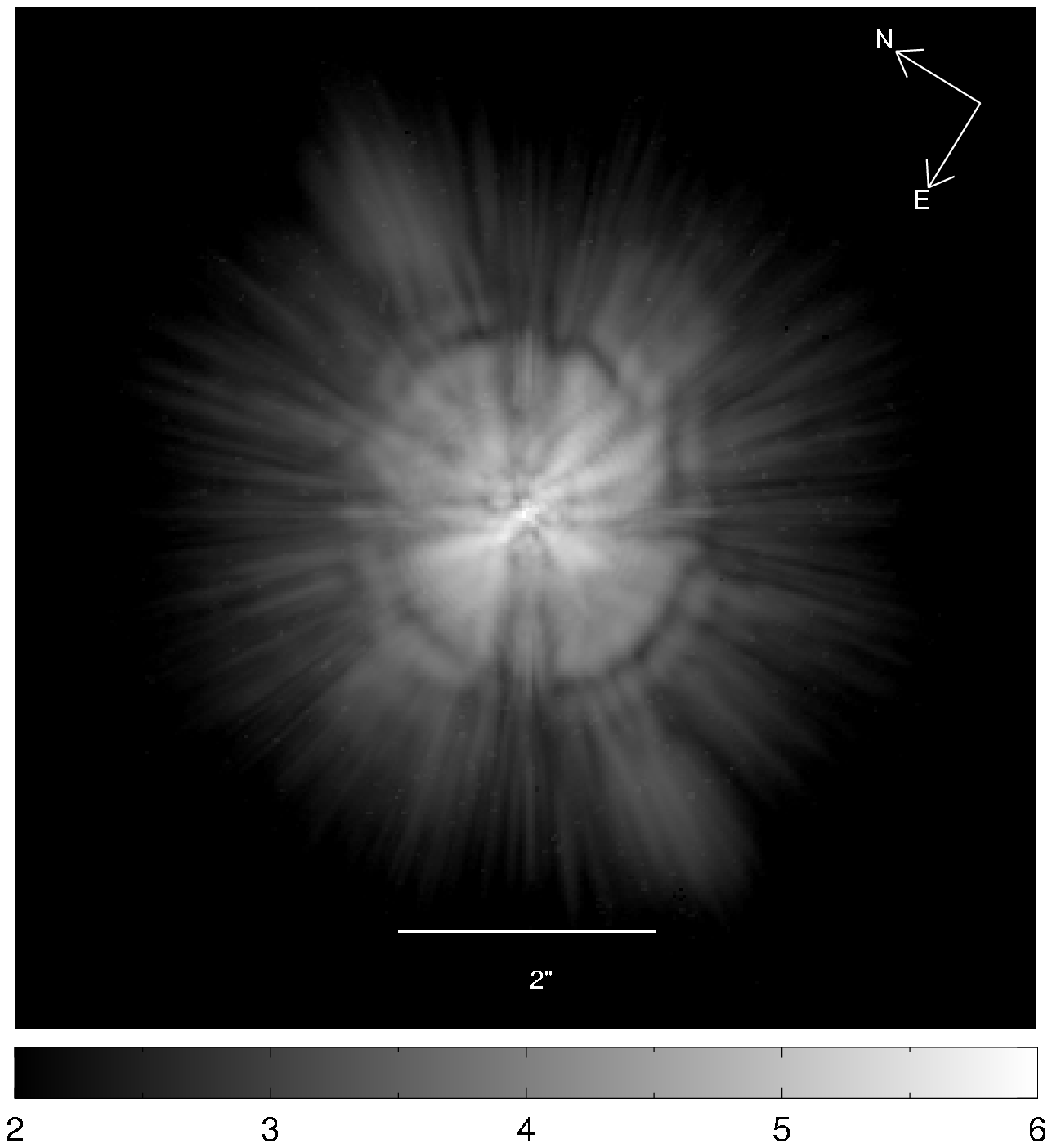}{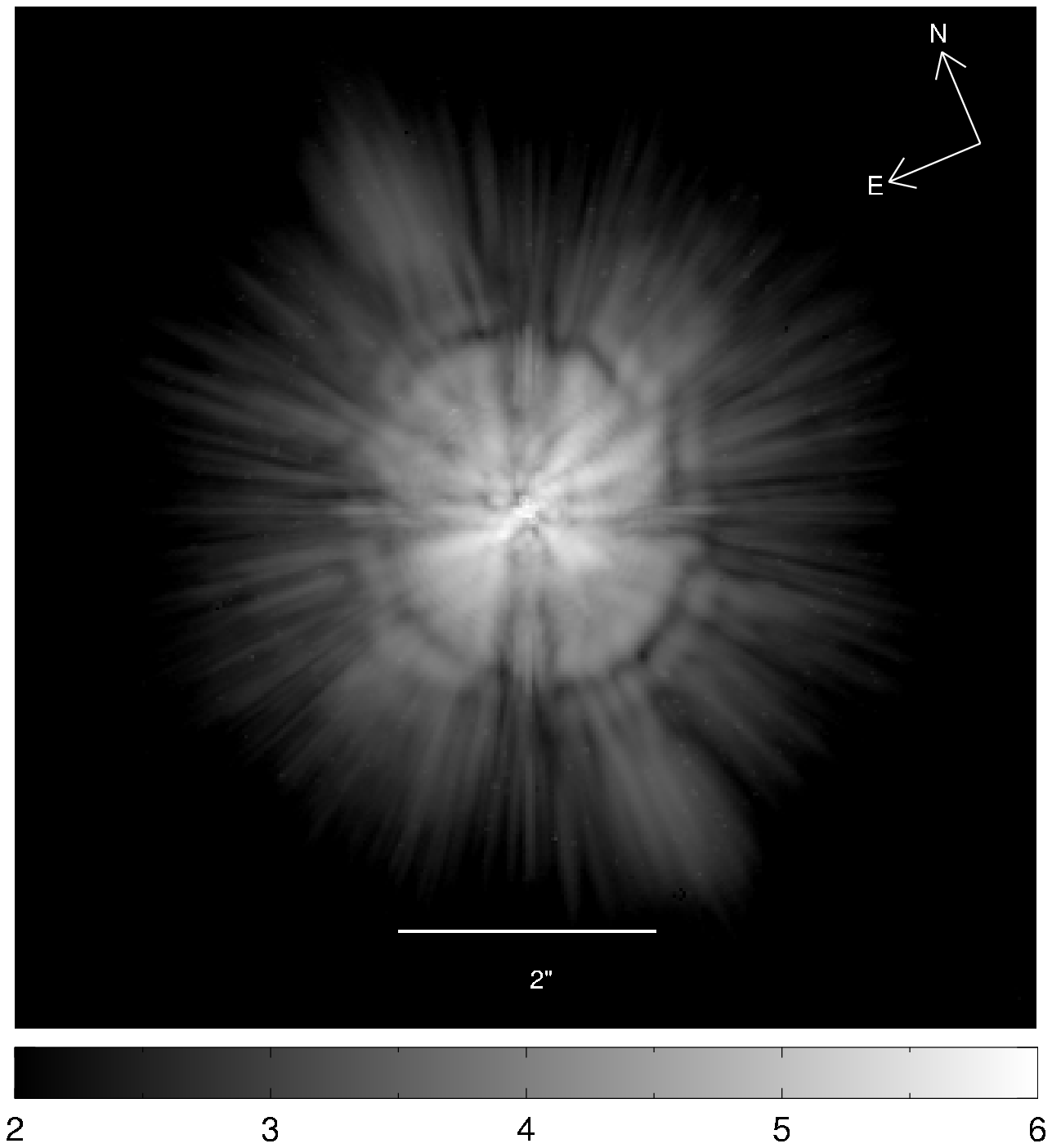}
\caption{Examples of unsubtracted occulted images of a disk candidate target (HD 97048, left)
                 and its corresponding PSF reference star (HD 80999, right). Circumstellar material is not visible prior to subtraction.  Shown with a logarithmic stretch.  Intensity units are in log(counts s$^{-1}$).\label{fig1}}
\end{figure}

\begin{figure}
\epsscale{.80}
\plotone{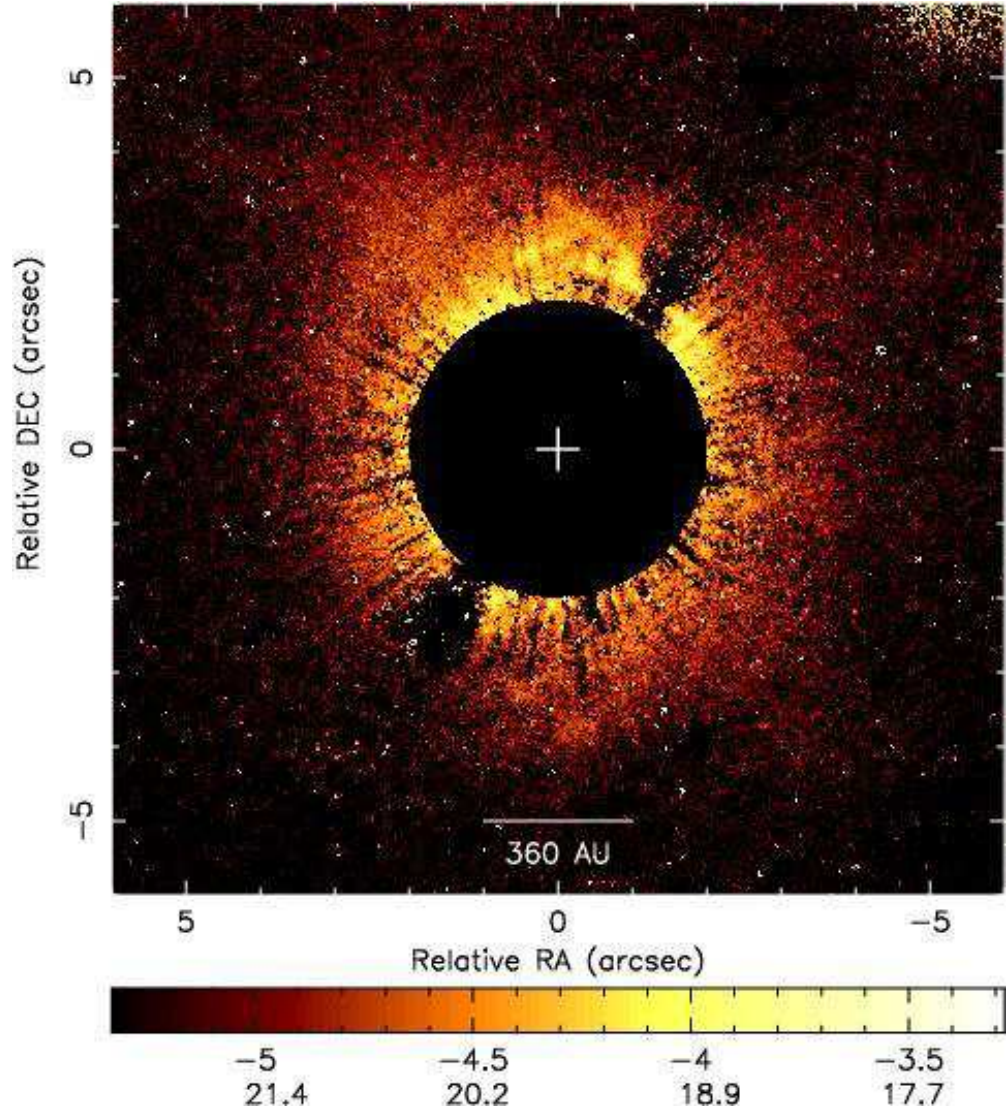}
\caption{{\it HST} ACS coronagraphic image of the circumstellar material around the Herbig Ae/Be star HD 97048 through the F606W filter. North is up and east is to the left.  The mask has a radius of 2\arcsec  and the plus sign indicates the position of the star.  Shown with a logarithmic intensity stretch.    Units on the intensity scale are Jy arcsec$^{-2}$ (top row) and mag arcsec$^{-2}$ (bottom row).  A sky value of 0.0171 counts s$^{-1}$ was used for the conversion.\label{fig2}}
\end{figure}

\clearpage

\begin{figure}
\epsscale{.80}
\plotone{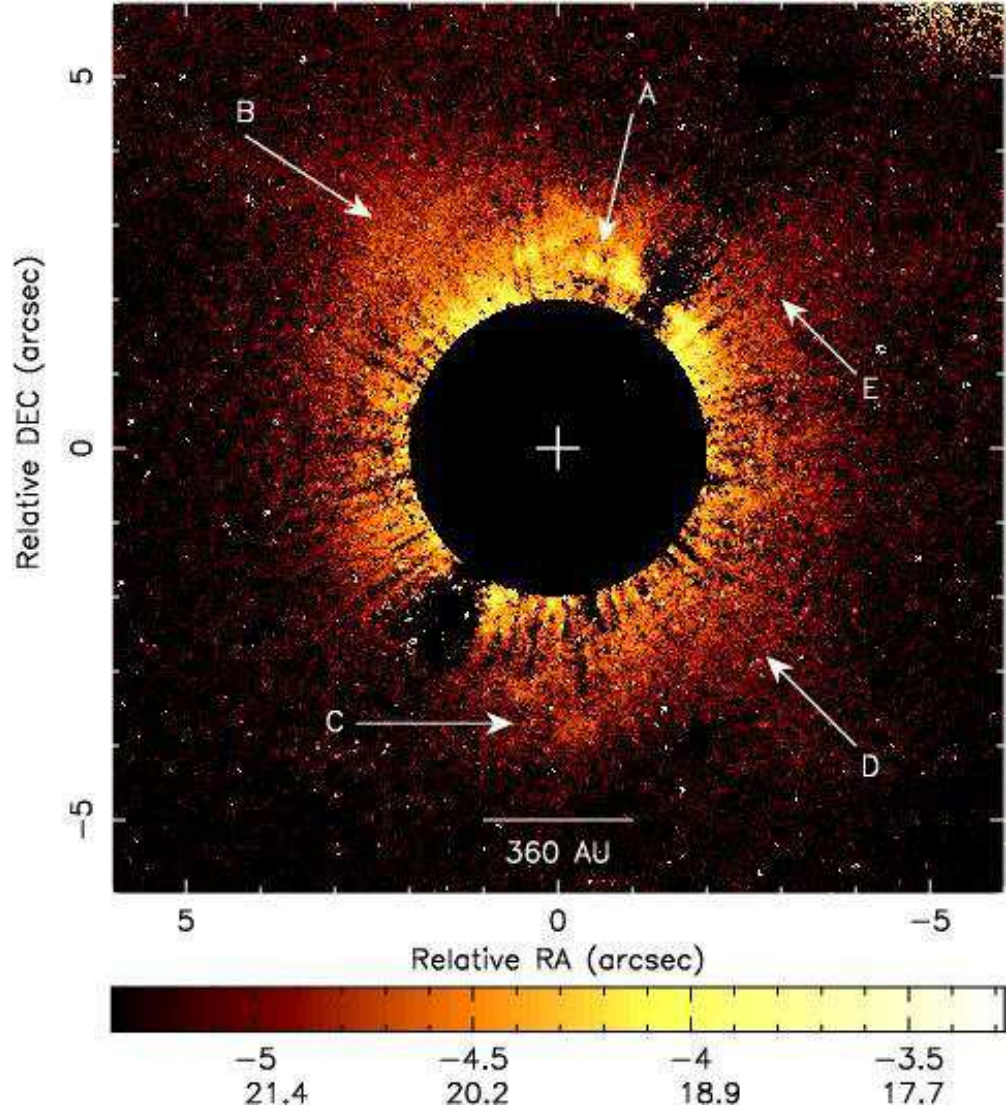}
\caption{The same as Figure 2 but with arrows included to identify observed structure.  The structure is labeled as follows: (A) northern spiral-like feature, (B) northeast extension, (C) southern clumps, (D) southwest extension, and (E) northwest extension.  S/N estimates for these features are given in the text.\label{fig3}}
\end{figure}

\clearpage

\begin{figure}
\epsscale{.80}
\plotone{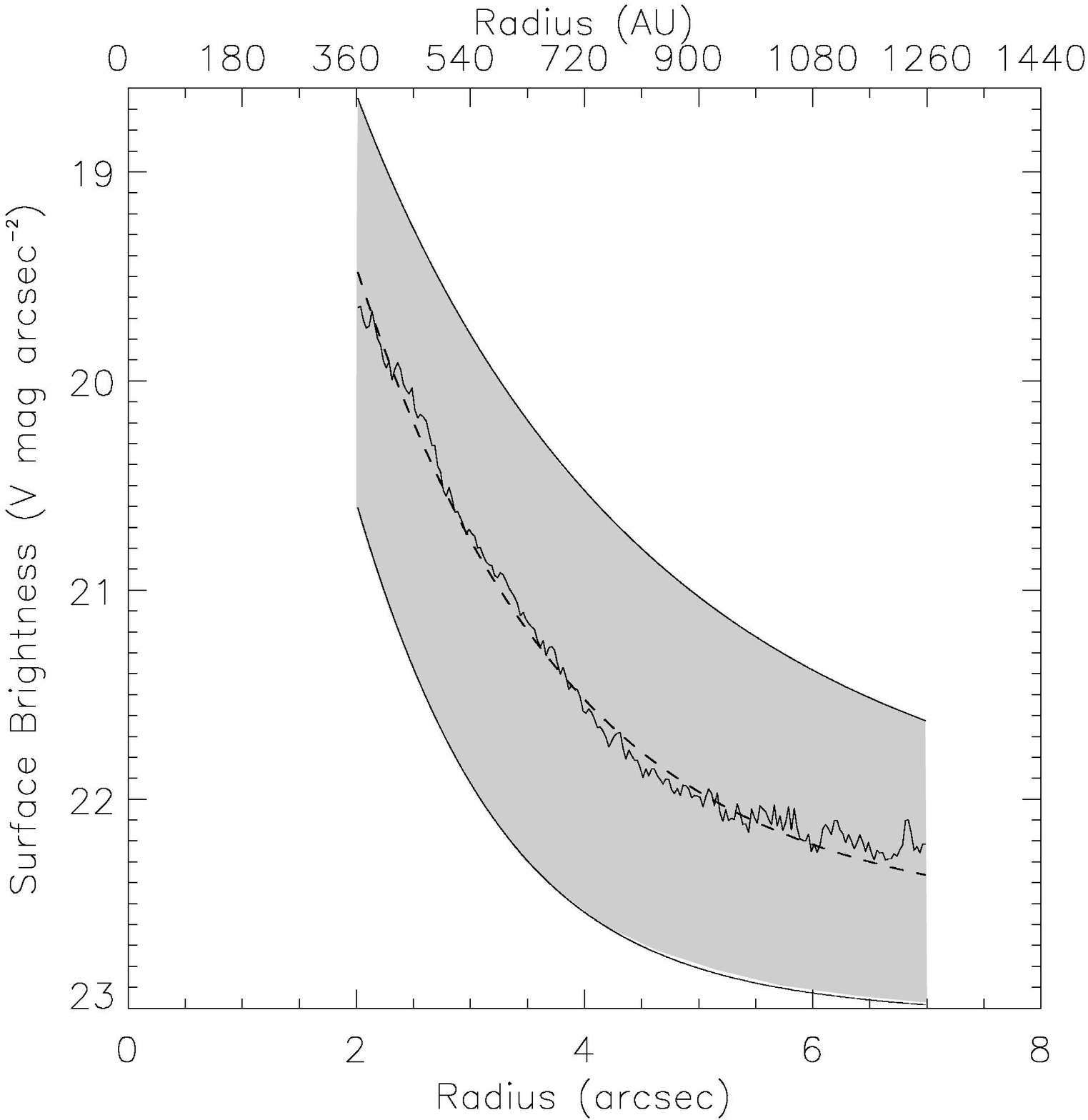}
\caption{$V$-band azimuthally-averaged radial surface brightness profile of the circumstellar material around HD 97048 from r $=$ 2$\arcsec$ to r $=$ 7$\arcsec$.  The dashed line indicates the best-fit to the measured data.  The shaded area shows the upper and lower bounds of the fit based on the uncertainties in the parameters.  This uncertainty reflects the standard deviation of the average surface brightness, which varies due to intrinsic structure (such as spiral-like features) and systematic effects from the PSF subtraction.\label{fig4}}
\end{figure}

\clearpage

\begin{figure}
\epsscale{.80}
\plotone{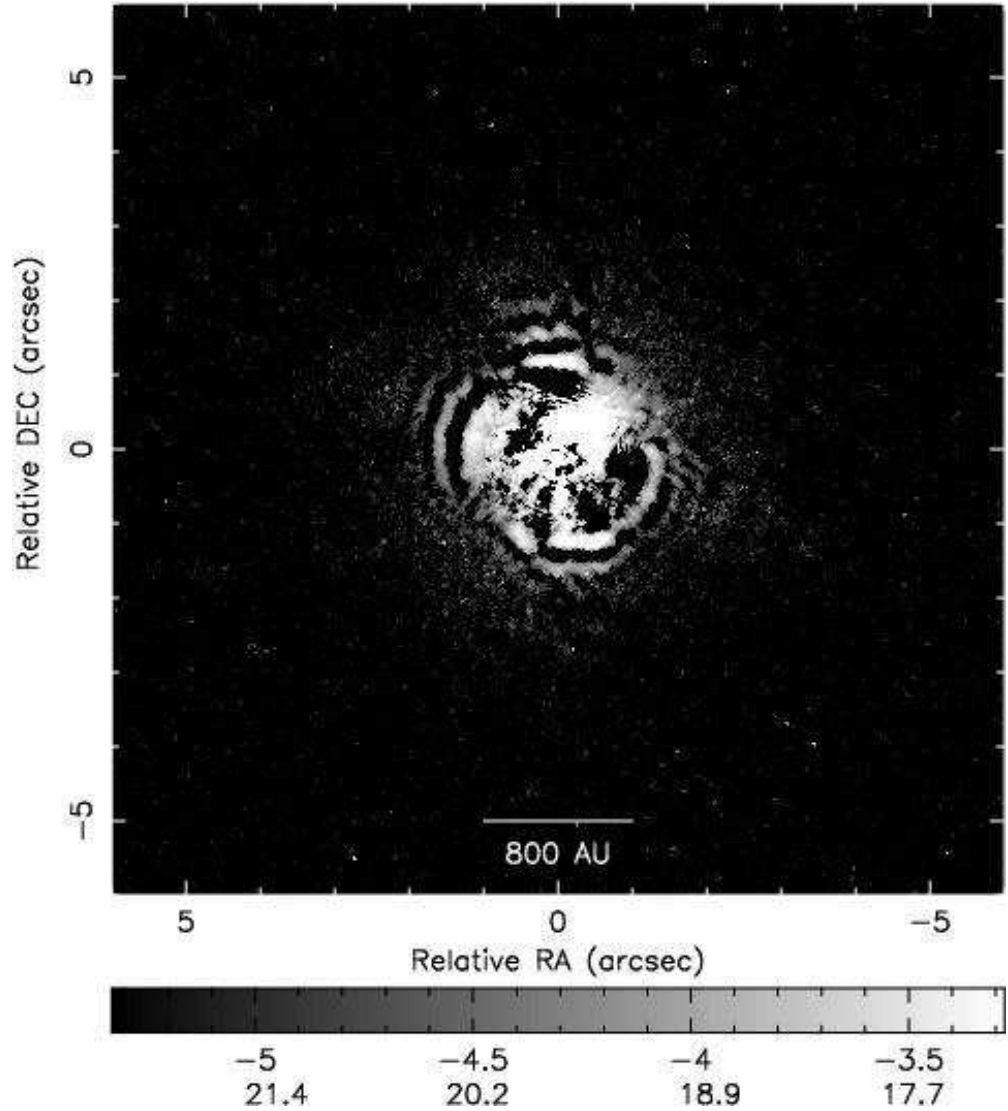}
\caption{The {\it HST}/ACS coronagraphic image of the Herbig Ae/Be star HD 34282 is an example of a non-detection.  North is up and east is to the left.  The 800 AU physical distance scale assumes the revised distance of 400 pc from \citet{pie03}.  Shown with a logarithmic stretch. Units on the intensity scale are Jy arcsec$^{-2}$ (top row) and mag arcsec$^{-2}$ (bottom row).  A sky value of 0.0064 counts s$^{-1}$ was used for the conversion.\label{fig5}}
\end{figure}

\begin{figure}
\epsscale{.80}
\plotone{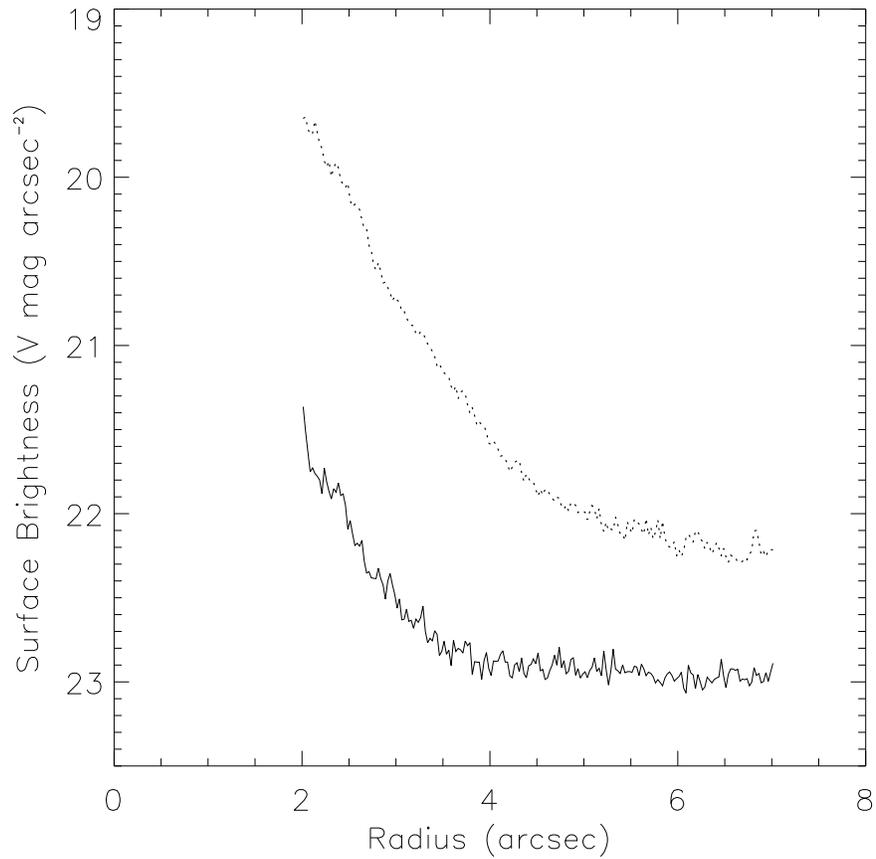}
\caption{$V$-band azimuthally-averaged radial surface brightness profile of the region surrounding the non-detection HD 34282 from r $=$ 2 -- 7$\arcsec$ (solid line). The profile for HD 97048 has been included for comparison (dotted line).  The scatter in the data points indicates the noise.\label{fig6}}
\end{figure}

\clearpage



\clearpage

\begin{deluxetable}{lcccccccccccc}
\tabletypesize{\scriptsize}
\footnotesize
\rotate
\tablecaption{Disk Candidate Star Properties\label{tbl1}}
\tablewidth{0pt}
\tablehead{ \colhead{Star} & \colhead{R.A.} & \colhead{Decl.} & \colhead{$V$} &
                    \colhead{\bv} & \colhead{Spec. Type} & \colhead{T$_{eff}$} & \colhead{d} & \colhead{L$_{IR}$/L$_{*}$}
                    & \colhead{Age} & \colhead{Circumstellar} & \colhead{Mid-IR} & \colhead{References}\\
                    &(2000) & (2000) & & & & (K) & (pc) & & (Myr) & Gas & Spectrum &}
\startdata
HD 34282 & 05 16 00.48 & -09 48 35.4 & 9.85 & 0.27 & A0 & 8625  & 400$^{+170a}_{-100}$  & 0.39 &  & CO &PAH &1, 2, 3, 4, 5, 6  \\

HD 97048 & 11 08 03.32 & -77 39 17.5 & 8.46 & 0.30 & A0 ep+sh & 10000  & 180$^{+30}_{-20}$  & 0.4 & $>$ 2   &[OI] & PAH & 1, 7, 8, 9, 10   \\

HD 139450 & 15 39 25.32 & -34 46 47.7 & 8.80 & 0.51 & G0/GI V & 5890  & 73$^{+7}_{-6}$  & $2.4 \times 10^{-3}$  & & Not Detected &  &1, 5, 11   \\

HD 158643 & 17 31 24.95 & -23 57 45.5 & 4.81 & 0.00 & A0 V & 10000  & 131$^{+17}_{-13}$  & $2.8 \times 10^{-2}$ & 0.3$^{+0.2}_{-0.1}$ & H$\alpha$, CO, CO$_{2}$, H$_{2}$O, NO &Silicates &  1, 7, 12, 13  \\

HD 159492 & 17 38 05.52 & -54 30 01.6 & 5.25 & 0.185 & A5 IV & 8200  & 42.2$^{+1.4}_{-1.3}$  & $1 \times 10^{-4}$ &  200   &  & & 1, 14  \\

HD 195627 & 20 35 34.85 & -60 34 54.3 & 4.759 & 0.272 & F0 V & 7000  & 27.6$^{+0.5}_{-0.5}$  & $1 \times 10^{-4}$ & 200  &  & & 1, 11, 14   \\
\enddata
\tablenotetext{a}{This is the revised distance for HD 34282 determined by \citet{pie03}.  Its {\it Hipparcos} distance is 160$^{+60}_{-40}$ pc \citep{van98}.}
\tablerefs{
(1) SIMBAD database; (2) Mer\'in et al. 2004; (3) Pi\'etu et al. 2003; (4) Sylvester et al. 1996; (5) Dent  et al. 2005; (6) Sloan et al. 2005; (7) van den Ancker et al. 1998;
(8) Van Kerckhoven et al. 2002; (9) Acke \& van den Ancker 2006; (10) van Boekel et al. 2004 (11) Sylvester \& Mannings 2000; (12) Jayawardhana et al. 2001; (13) van den Ancker et al. 2001;
(14) Zuckerman \& Song 2004.}
\end{deluxetable}

\clearpage

\begin{deluxetable}{lccccc}
\footnotesize
\tablecaption{PSF Reference Star Properties$^{a}$\label{tbl2}}
\tablewidth{0pt}
\tablehead{ \colhead{PSF Ref. Star} & \colhead{R.A. (2000)} & \colhead{Decl. (2000)} & \colhead{$V$} & 
                     \colhead{\bv} & \colhead{Spec. Type}}
\startdata
HD 36863 & 05 34 34.34 & -01 44 37.4 & 8.28 & 0.26 & A0   \\
HD 80999 & 09 20 54.04 & -55 48 43.5 & 7.98 & 0.31 & A0 II  \\
HD 144766 & 16 08 07.12 & -18 14 35.3 & 7.03 & 0.52 & G1 V  \\
HD 159217 & 17 35 39.59 & -46 30 20.5 & 4.571 & -0.019 & A0 V  \\
HD 135379 & 15 17 30.85 & -58 48 04.3 & 4.169 & 0.10 & A3 V  \\
HD 219571 & 23 17 25.77 & -58 14 08.6 & 4.369 & 0.366 & F1 III  \\
\enddata
\tablenotetext{a}{Stellar properties collected from the SIMBAD database.}
\end{deluxetable}

\clearpage

\begin{deluxetable}{lcccccc}
\tabletypesize{\scriptsize}
\footnotesize
\tablecaption{Observational Parameters for {\it HST} Program 10425\label{tbl3}}
\tablewidth{0pt}
\tablehead{ \colhead{Star} & \colhead{Target Type} & \colhead{Date} & \colhead{Exposure} & 
                     \colhead{CR-split} & \colhead{Gain} & \colhead{Exp. Type$^a$}}
\startdata
HD 34282    & Herbig Ae/Be   & 2005 Jan 10    & 3 x 0.1s         & None    & 4    & D   \\
                      &                    &                            & 1 x 6s            & 3            & 4    & D   \\
                      &                    &                            & 1 x 200s       & 3            & 2    & C   \\
                      &                    &                            & 1 x 1780s     & 3           & 2     & C   \\
HD 36863   & PSF Ref.   & 2005 Jan 10    & 1 x 200s       & 3           & 2      &  C       \\
                    &                  &                        & 1 x 1480s & 3 & 2 & C      \\
                    &                  &                        & 3 x 0.1s  & None & 4 & D  \\
                    &                  &                        & 1 x 3s   & 3  & 4 & D           \\
HD 139450 & Debris Disk & 2005 Jan 14 & 3 x 0.1s & None & 4 & D  \\
                      &                 &                        & 1 x 3s & 3 & 4 & D             \\
                      &                 &                        & 1 x 200s & 3 & 2 & C        \\
                      &                 &                        & 1 x 1830 & 3 & 2 & C        \\
HD 144766 & PSF Ref. & 2005 Jan 14 & 1 x 200s & 3 & 2 & C      \\
                      &                  &                         & 1 x 1480s & 3 & 2 & C    \\
                      &                  &                         & 3 x 0.1s & None & 4 & D  \\
                      &                  &                         & 1 x 1s & 3 & 4 & D            \\
HD 195627 & Debris Disk & 2005 Jun 21 & 3 x 0.1s & None & 4 & D  \\
                      &                 &                        & 1 x 96s & 3 & 2 & C          \\
                      &                 &                        & 1 x 2135s & 7 & 2 & C     \\
HD 219571 & PSF Ref. & 2005 Jun 21 & 1 x 12s & 3 & 2 & C        \\
                      &                  &                         & 1 x 2205s & 7 & 2 & C   \\
                      &                  &                         & 3 x 0.1s & None & 4 & D \\
HD 159492 & Debris Disk & 2005 Jul 1 & 3 x 0.1s & None & 4 & D      \\
                      &                &                      & 1 x 150s & 3 & 2 & C            \\
                      &                &                      & 1 x 2016s & 7 & 2 & C         \\
HD 135379 & PSF Ref. & 2005 Jul 1 & 1 x 12s & 3 & 2 & C              \\
                      &                  &                     & 1 x 2205s & 7 & 2 & C         \\
                      &                  &                     & 3 x 0.1s & None & 4 & D    \\ 
HD 97048 & Herbig Ae/Be & 2005 Jul 16 & 3 x 0.1s & None & 4 & D    \\
                    &                &                       & 1 x 3s & 3 & 4 & D                \\
                    &                &                       & 1 x 200s & 3 & 2 & C           \\
                    &                &                       & 1 x 2200s & 3 & 2 & C         \\
HD 80999 & PSF Ref. & 2005 Jul 16 & 1 x 200s & 3 & 2 & C         \\
                    &                  &                       & 1 x 1695s & 3 & 2 & C       \\
                    &                  &                       & 3 x 0.1s & None & 4 & D   \\
                    &                  &                       & 1 x 3s & 3 & 4 & D             \\
HD 158643 & Herbig Ae/Be & 2005 Aug 14 & 3 x 0.1s & None & 4 & D \\
                      &                 &                         & 1 x 96s & 3 & 2 & C         \\
                      &                 &                         & 1 x 1920s & 6 & 2 & C     \\
HD 159217 & PSF Ref. & 2005 Aug 14 & 1 x 18s & 3 & 2 & C        \\
                      &                  &                         & 1 x 2086s & 7 & 2 & C    \\
                      &                  &                         & 3 x 0.1s & None & 4 & D \\
\enddata
\tablenotetext{a}{D $=$ direct exposure (used for photometric measurements), and C $=$
                               coronagraphic exposure.}
\end{deluxetable}

\clearpage

\begin{deluxetable}{ccc}
\footnotesize
\tablecaption{Disk Candidate and PSF Reference Star $V$-band Photometry\label{tbl4}}
\tablewidth{0pt}
\tablehead{ \colhead{Star} & \colhead{Type} & \colhead{Magnitude$^{a}$}}
\startdata
HD 34282  &  Herbig Ae/Be   &  9.94 $\pm$ 0.03  \\
HD 36863  &  PSF Ref.   &  8.18 $\pm$ 0.03  \\
HD 97048  &  Herbig Ae/Be   &  8.36 $\pm$ 0.03  \\
HD 80999  &  PSF Ref.   &  7.85 $\pm$ 0.03  \\
HD 139450  &  Debris Disk  &  8.72 $\pm$ 0.03  \\
HD 144766  &  PSF Ref.  &  7.04 $\pm$ 0.03  \\
HD 158643  &  Herbig  Ae/Be &  4.80 $\pm$ 0.03  \\
HD 159217  &  PSF Ref.  &  4.58 $\pm$ 0.03  \\
HD 159492  &  Debris Disk  &  5.23 $\pm$ 0.03  \\
HD 135379  &  PSF Ref.  &  4.07 $\pm$ 0.03  \\
HD 195627  &  Debris Disk  &  4.75 $\pm$ 0.03  \\
HD 219571  &  PSF Ref.  &  3.97 $\pm$ 0.03  \\
\enddata
\tablenotetext{a}{The errors include an estimate of the uncertainty in the magnitude system transformation.}
\end{deluxetable}

\clearpage

\begin{deluxetable}{lr}
\tablecaption{HD 97048 Circumstellar Material Properties\label{tbl5}}
\tablewidth{0pt}
\tablehead{ \colhead{Property} & \colhead{Value}}
\startdata
Outer Radius & $\ga$ 4$\arcsec$ (720 AU)  \\
Peak $V$-band Surface Brightness & 19.6 $\pm$ 0.2 mag arcsec$^{-2}$  \\
Surface Brightness Radial Falloff & {\it r}$^{-3.3 \pm 0.5}$  \\
Integrated Flux & 16.8 $\pm$ 0.1 mag  \\
L$_{sca}$/L$^{a}_{*}$ &  $>$ $8.4 \times 10^{-4}$   \\
\enddata
\tablenotetext{a}{This ratio is a lower limit since the disk is optically thick.}
\end{deluxetable}

\clearpage

\begin{deluxetable}{lcc}
\tabletypesize{\scriptsize}
\tablecaption{Non-detection Surface Brightness Limits\label{tbl6}}
\tablewidth{0pt}
\tablehead{ \colhead{Star} & \colhead{Surface Brightness 3$\sigma$ Upper Limit} 
	& \colhead{Sky Brightness 3$\sigma$ Upper Limit}\\
	& (mag arcsec$^{-2}$)   &  (mag arcsec$^{-2}$)}
\startdata
HD 34282   &    20.5   &   21.2   \\
HD 139450 &    19.2   &   21.8   \\
HD 158643 &    15.3   &   19.5  \\
HD 159492 &    16.0   &   20.4   \\
HD 195627 &    15.4   &   20.1   \\
\enddata
\end{deluxetable}

\clearpage

\begin{deluxetable}{lcccccccc}
\tabletypesize{\scriptsize}
\rotate
\tablecaption{Properties of a Subset of Herbig Ae/Be Stars Imaged with {\it HST} Coronagraphs\label{tbl7}}
\tablewidth{0pt}
\tablehead{ \colhead{Star} & \colhead{R$_{out}$} & \colhead{d} & \colhead{Spec. Type} &
                    \colhead{Age} & \colhead{L$_{IR}$/L$_{*}$} & \colhead{Radial SB Power-Law} & \colhead{{\it HST} Instrument} & \colhead{References}\\
                    &  (AU) & (pc)  &   &  (Myr)  &  &   &  }
\startdata
HD 100546$^{a}$    & 515  & 103$^{+7}_{-6}$  & B9 Vne & $\geq$ 10 & 0.51 & {\it r}$^{-3.1}$ (within 5$\arcsec$) & ACS, NICMOS, STIS & 1, 2, 3, 4, 5   \\

AB Aurigae    & 1300  & 144$^{+23}_{-17}$  & A0 Ve+sh    & 2$^{+1.16}_{-0.75}$ &   0.48 &   {\it r}$^{-2}$        & STIS & 4, 5, 6 \\

HD 163296    & 450    & 122$^{+17}_{-13}$  & A1 Ve    & 4$^{+6}_{-2.5}$  &  0.16 &  {\it r}$^{-3.5}$ ({\it r} $\ge$ 370 AU) & STIS &  4, 5, 7  \\

HD 141569A & 1200  &   99 $^{+9}_{-8}$  & B9.5 Ve & 5 $\pm$ 3 & $8.4 \times 10^{-3}$ & {\it r}$^{-3.2}$ (190 $<$ {\it r} $<$ 250 AU)  & ACS, NICMOS, STIS & 4, 8, 9, 10, 11, 12, 13  \\

HD 97048      & 720    & 180$^{+30}_{-20}$  & A0 ep+sh & $>$ 2    &  0.4   &{\it r}$^{-3.3}$ & ACS & 4, 14, 15, 16 \\

HD 34282 &  Not Detected & 400$^{+170}_{-100}$ & A0  &   &  0.39 &Not Detected &  ACS & 4, 13, 14, 15, 17 \\

HD 158643  & Not Detected &  131$^{+17}_{-13}$ & A0 V & 0.3$^{+0.2}_{-0.1}$  & $2.8 \times 10^{-2}$ & Not Detected &  ACS &  4, 14, 15, 18 \\
\enddata
\tablenotetext{a}{The R$_{out}$ and power-law values given in the table for HD 100546 describe its disk.  The system's envelope extends to approximately 1000 AU with a {\it r}$^{-2.2}$ drop off.}
\tablerefs{
(1) Grady et al. 2001; (2) Augereau et al. 2001; (3) Ardila  et al. 2005; (4) van den Ancker et al. 1998;
(5) Meeus et al. 2001; (6) Grady et al. 1999; (7) Grady et al. 2000; (8) Clampin et al. 2003;
(9) Weinberger et al. 1999; (10) Weinberger et al. 2000; (11) Augereau et al. 1999; (12) Mouillet et al. 2001; (13) Sylvester et al. 1996;
(14) this work; (15) SIMBAD database; (16) Van Kerckhoven et al. 2002; (17) Pi\'etu et al. 2003; (18)Jayawardhana et al. 2001.}
\end{deluxetable}

\clearpage

\begin{deluxetable}{lccccccc}
\tabletypesize{\scriptsize}
\tablecaption{Properties of a Subset of Debris Disk Stars Imaged with the ACS Coronagraph\label{tbl8}}
\tablewidth{0pt}
\tablehead{ \colhead{Star} & \colhead{R$_{out}$} & \colhead{d} & \colhead{Spec. Type} &
                    \colhead{Age} & \colhead{L$_{IR}$/L$_{*}$} & \colhead{Disk Surface Brightness$^{a}$} & \colhead{References}  \\
                    &  (AU)  & (pc)  &    & (Myr)  &      & (mag arcsec$^{-2}$) &}
\startdata
Fomalhaut & 158 & 7.7 & A3 V &  200 $\pm$ 100 & $5 \times 10^{-5}$ & 20.6 $\pm$ 0.1 & 1, 2, 3  \\
HD 53143  & $>$ 110 & 18.4 & K1 V & 1000 $\pm$ 200 & $2.5 \times 10^{-4}$ & 22.0 $\pm$ 0.3  & 4, 5    \\
HD 107146 & $>$ 185 & 28.4 & G2 V & 30 -- 250 &  $1.2 \times 10^{-3}$ &   & 6, 7  \\
HD 139664 & 109 & 18.5 & F5 V &  300$^{+700}_{-200}$ & $0.9 \times 10^{-4}$ & 20.5 $\pm$ 0.3  &   4, 5  \\
HD 139450  &  Not Detected & 73$^{+7}_{-6}$ & G0/GI V &  &  $2.4 \times 10^{-3}$ & $>$ 19.2 & 8, 9, 10, 11  \\
HD 159492  &  Not Detected & 42.2$^{+1.4}_{-1.3}$ & A5 IV & 200 & $1 \times 10^{-4}$ & $>$ 16.0  & 5, 8, 10  \\
HD 195627  &  Not Detected & 27.6$^{+0.5}_{-0.5}$ & F0 V & 200 & $1 \times 10^{-4}$ & $>$ 15.4 &  5, 8, 9, 10 \\
\enddata
\tablenotetext{a}{The Fomalhaut surface brightness is taken from quadrant one (Q1) in a combined F606W and F814W image as described by Kalas et al. 2005. The surface brightnesses for HD 53143 and HD 139664 are peak values.  The $V$-band 3$\sigma$ upper limits are given for our non-detections.}
\tablerefs{
(1) Kalas et al. 2005; (2) Barrado y Navascu\'es 1998; (3) Decin et al. 2003; (4) Kalas  et al. 2006; (5) Zuckerman \& Song 2004; (6) Ardila et al. 2004; (7) Williams et al. 2004; (8) this work; (9) Sylvester \& Mannings 2000; (10) SIMBAD database; (11) Dent et al. 2005;.}
\end{deluxetable}



\end{document}